# Enhancement in optically induced ultrafast THz response of MoSe$_2$MoS$_2$ heterobilayer

Sunil Kumar,[1,*] Arvind Singh,[1] Sandeep Kumar,[1] Anand Nivedan,[1] Marc Tondusson,[2] Jerome Degert,[2] Jean Oberle,[2] Seok Joon Yun,[3] Young Hee Lee,[3] and Eric Freysz[2,*]

[1]*Femtosecond Spectroscopy and Nonlinear Photonics Laboratory,
Department of Physics, Indian Institute of Technology Delhi, New Delhi 110016, India*
[2]*Univ. Bordeaux, CNRS, LOMA UMR 5798, 33405 Talence, France*
[3]*Center for Integrated Nanostructure Physics (CINAP),
Institute for Basic Science (IBS), Sungkyunkwan University, Suwon 16419, Republic of Korea*
[*]eric.freysz@u-bordeaux.fr
[*]kumarsunil@physics.iitd.ac.in

**THz conductivity of large area MoS$_2$ and MoSe$_2$ monolayers as well as their vertical heterostructure, MoSe$_2$MoS$_2$ is measured in the 0.3-5 THz frequency range. Compared to the monolayers, the ultrafast THz reflectivity of the MoSe$_2$MoS$_2$ heterobilayer is enhanced many folds when optically excited above the direct band gap energies of the constituting monolayers. The free carriers generated in the heterobilayer evolve with the characteristic times found in each of the two monolayers. Surprisingly, the same enhancement is recorded in the ultrafst THz reflectivity of the heterobilayer when excited below the MoS$_2$ bandgap energy. A mechanism accounting for these observations is proposed.**

The family of two-dimensional semiconducting transition metal dichalcogenides (2D-TMDs) such as MoS$_2$, WS$_2$, MoSe$_2$ and so on, has grown significantly [1] and it is likely to remain one of the leading topics in science for many years to come due to many facets of scientific findings and knowledge they can contribute to. Heterostructures of such 2D systems offer not only a way to study their electronic properties and interlayer interactions but also provide a rich playground to expand the physics of the constituting layers [2], eventually to open enormous possibilities of utilizing them for various technological applications [3,4]. Unlike conventional semiconductor heterostructures, van der Waals (vdW) 2D heterostructures having atomically sharp interfaces are relatively easy to make either via mechanical exfoliation from bulk crystals by scotch tape method [5] or via bottom-up approaches such as using chemical vapor deposition techniques [6–8], without considering the lattice mismatch between different 2D layers. Heterostructures formed by two TMD monolayers can exhibit type-I or type-II electronic band alignments [9,10], leading to the formation of interlayer radiative excitons where the bottom of the conduction band and the top of the valence band reside in different layers. In such a case, valley lifetimes become longer than that for intralayer excitons [11,12]. More importantly, the electronic band alignment provides spatial separation of electrons and holes after photoexcitation in the heterostructures, facilitating an important virtue for their applications in photovoltaic devices [13,14]. The interfacial electronic interactions and charge transfer/separation between constituting TMDs is one of the keys to determine the characteristics of such heterostructures and hence the performance of devices made from them.

In a few reports on vertically stacked vdW heterostructures of TMDs, possible mechanisms for the charge transfer have been inferred from time-resolved optical absorption and photoluminescence studies [15–20]. Photoluminescence time-resolved spectroscopy has been used to study the fast interlayer energy transfer in MoSe$_2$/WS$_2$ heterostructures [21], while ultrafast time-resolved optical pump probe spectroscopy has been used to study possible charge transfer mechanisms, including the exciton localization [22], interlayer hot excitons formation [18], and coherent charge transfer [23]. Note that all the above studies are usually done at optical photon energies above the bandgap energies of the corresponding semiconducting layers. Terahertz (THz) spectroscopy using sub-picosecond electrical transients, which is an ideal tool to study the low energy electronic processes, especially due to its sensitivity to the electrical conductivity in relevant materials for ultrafast optoelectronics, has been rarely used [24]. In fact, ultrafast THz pulses can help in investigating the time- and frequency-resolved hot charge carrier density, mobility, and photoconductivity within the heterostructure, the layers and the interfaces, in a noncontact manner to reveal their true characteristics. The THz conductivity can be modulated by modulating the free carrier density. Optical excitation by appropriate light is the most convenient way to achieve this. Moreover, the use of an ultrafast optical pulse for generating photocarriers can also provide a handle on the THz conductivity modulation in real time where the characteristic hot carrier relaxation times are underplaying. This is the prime matter of concern in our paper here.

By using sub-200 femtosecond THz pulses, we measure electrical conductivity of MoS$_2$ and MoSe$_2$ monolayers, and MoSe$_2$MoS$_2$ vertical heterostructure. Both the MoS$_2$ and MoSe$_2$ monolayers are direct bandgap semiconductors which form a vdW heterostructure with type–II electronic band arrangement. The static THz conductivity of the MoS$_2$ monolayer is nearly zero while it is finite for the MoSe$_2$ monolayer in the entire frequency range of 0.2-5 THz. For the MoSe$_2$MoS$_2$ heterostructure, the THz conductivity is found to be nearly equal to the arithmetic mean of the conductivities of the two constituting layers. In sharp contrast, the dynamic THz response from THz reflectivity measurements of the optically excited MoSe$_2$MoS$_2$ is found to be at least an order of magnitude higher than either of the two constituting monolayers. Much importantly, by creating photocarriers only in the MoSe$_2$ layer, similar enhancement in THz response of the heterostructure is observed. It is due to the almost instantaneous movement of photoelectrons across the atomistic interface between the MoSe$_2$ and MoS$_2$. From the saturation of the dynamic THz response of MoSe$_2$MoS$_2$ at excitation photon energy of 1.55 eV, i.e., below the MoS$_2$-bandgap energy, we find that about 90% of the photoelectrons transiently transferred from MoSe$_2$ to MoS$_2$ layer.

*Materials:* Large area (~15 mm x 15 mm) monolayers of MoSe$_2$ and MoS$_2$ were grown on SiO$_2$/Si substrates by two-zone furnace atmospheric pressure chemical vapor deposition technique [25]. These were then transferred onto quartz substrates using poly(methyl methacrylate) assisted wet-etching method. The MoSe$_2$MoS$_2$-heterobilayer was fabricated by transferring MoSe$_2$ layer onto the MoS$_2$ sample. The lateral size of the almost uniform monolayers of MoS$_2$ and MoSe$_2$ and their heterobilayer, MoSe$_2$MoS$_2$ on the substrates was sufficiently large for carrying out the THz experiments discussed in this paper (see supplementary information). The quality of the samples was also assessed using various techniques including Raman spectroscopy [26] and UV-Visible absorption spectroscopy.

*Absorption and Raman spectroscopy:* The respective optical absorption spectra of the TMD layers also exhibit the characteristic A-, B- and C- excitonic features, which remain intact in the MoSe$_2$MoS$_2$-heterobilayer (see Fig. S1 in the supplementary information). Raman spectroscopy is a tool of choice to characterize 2D-TMDs. The Raman spectra were recorded using a Thermo-Scientific DXR Raman Microscope by exciting the samples at 532 nm laser wavelength. Characteristic Raman frequencies, corresponding to the symmetric in-plane $E_{2g}^1$ phonon and out-of-plane A$_{1g}$ phonon modes of both the MoS$_2$ and MoSe$_2$ as indicated in Fig. 1(a) for all the three samples, are in excellent agreement with the reported values in the literature [27–29]. It can be seen from Fig. 1(a) that the $E_{2g}^1$ modes of both the monolayers (288 cm$^{-1}$ for MoSe$_2$ and 386 cm$^{-1}$ for MoS$_2$) are weakly softened by ~1 cm$^{-1}$ in the MoSe$_2$MoS$_2$-heterobilayer. However, the shift in the frequency of the A$_{1g}$ mode in the latter is opposite for the MoS$_2$ and MoSe$_2$ monolayers, i.e., for MoSe$_2$ (242 cm$^{-1}$) it softens by ~1 cm$^{-1}$, while for MoS$_2$ (404 cm$^{-1}$) it gets hardened by a similar number. This opposite shift in the frequency of the out of plane A$_{1g}$ mode of the two monolayers is due to the fact that MoS$_2$ is mechanically supported from both the sides, i.e., by MoSe$_2$ on one side and the substrate on the other. This type of opposite shift in the Raman frequencies has also been reported in the literature for MoS$_2$ and WS$_2$MoS$_2$ [29,30].

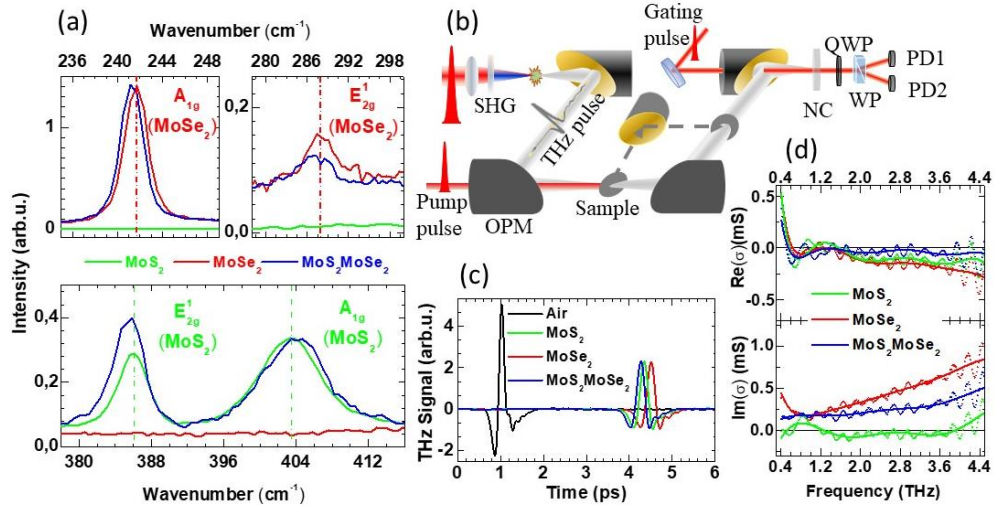

**FIGURE 1.** Characterization and optical/THz experiments on MoS$_2$ monolayer, MoSe$_2$ monolayer and the MoSe$_2$MoS$_2$-heterobilayer, all supported on quartz substrates. **(a)** Raman spectra: vertical dashed lines to mark the positions of Raman active modes, $E_{2g}^1$ and A$_{1g}$, in the monolayers and the heterobilayer. **(b)** Schematic of the experimental setup used for THz time-domain spectroscopy and optical pump-THz probe time-resolved spectroscopy in both the reflection and transmission configurations. SHG: second harmonic generation crystal, OPM: off-axis parabolic mirror, NC: nonlinear crystal for electro-optic sampling, QWP: quarter wave plate, WP: Wollaston prism, PD: photodiode. **(c)** Transmission THz time-domain traces taken in air (without any sample), MoS$_2$, MoSe$_2$ monolayers and the MoSe$_2$MoS$_2$-heterobilayer. **(d)** Real (Re($\sigma$)) and imaginary (Im($\sigma$)) parts of the static sheet conductivity ($\sigma$). Continuous lines represent the mean obtained using a polynomial fitting.

***Static THz spectroscopy:*** Part of the experimental setup used for the measurement of static and dynamic THz response of our materials is shown in Fig. 1(b). Terahertz pulses of sub-200 fs duration were generated in air plasma and detected in a nonlinear crystal (NC) by an electro-optic sampling technique in the usual manner [31], as shown in Fig. 1(b). For driving the THz setup, femtosecond pulses with time-duration of ~50 fs were taken from a Ti:Sapphire based regenerative amplifier operating at 1 kHz repetition rate and a central wavelength (photon energy) of 800 nm (~1.55 eV). The main part of the 800 nm laser output was used to create a plasma in air by focusing the fundamental and its second harmonic from a BBO crystal, using a common biconvex lens with a 150 mm focal length. The optical beam was separated from the produced THz pulses using a highly resistive silicon wafer. A much weaker part of the femtosecond laser beam was used as sampling/gating beam for detecting/recording the THz pulses in a NC through electro-optic effect. From the source point till the detection at the nonlinear crystal, which was a 200 µm thick <110> GaP stuck on a 3 mm thick <100> GaP substrate, four silver-coated off-axis parabolic mirrors (OPMs) having focal length of 150 mm were used in a 8f imaging geometry, with the sample placed at the center between the two inner parabolic mirrors. The THz beam and gating beam are collinearly and temporally matched on the GaP crystal. Then, a combination of a quarter wave plate (QWP), a Wollaston prism (WP) and balanced photodiodes (PDs) ensures the measurements of an electrical signal on a lock-in amplifier that is proportional to the magnitude and phase of the THz pulse falling on the GaP crystal. The complete THz setup was enclosed in a box under continuous purging of dry air to minimize THz losses from the generation point in the air-plasma till the detection point at the GaP crystal. With a few adjustments, our experimental setup allows measurements of the static and dynamic THz response.

THz time-domain scans were recorded in air, i.e., without any sample, on quartz substrate for reference and on the two monolayers and the their heterobilayer, all supported on quartz substrates. The effect of the quartz substrate was appropriately removed from the overall response of the layer samples on quartz substrates by performing reference experiments on the bare quartz plate from the same batch of substrates. Our THz experimental setup allows measurements in both the reflection and transmission modes. In each case, reference signals were also taken on the quartz substrate alone. For reflection measurements, one additional parabolic mirror and a large silver-coated plane mirror were used at f-distances as shown in the schematic of Fig. 1(b). The index of refraction $n(\Omega)$ and extinction coefficient $\kappa(\Omega)$ of the quartz substrate were determined using the standard procedure in THz time-domain spectroscopy (Fig. S3 in the supplementary information). These two parameters were used in combination with the time-domain complex electric field waveforms, $E_{THz}(t)$ as shown in Fig. 1(c) to analyze the static THz conductivity of the layers. The small temporal shift among the THz scans for the various thin samples displayed in Fig. 1(c), arises from the slight difference in the thicknesses of the quartz substrates used for each of them and checked independently. The thicknesses of the quartz substrates were 1.0444±0.0001 mm, 1.0860±0.0001 mm and 1.0153±0.0001 mm for $MoS_2$, $MoSe_2$ and $MoSe_2MoS_2$ samples, respectively. In fact, the mean value of the THz refractive index (n~1.9613) of the quartz substrate obtained from the temporal shifts in Fig. 1(c) and corresponding thicknesses are in perfect agreement. Though the setup in Fig. 1(b) allowed large bandwidth of ~0.2-8 THz, the losses in the quartz substrate limit the reliable evaluation of static THz conductivities of our TMD samples to ~0.2-5 THz.

The absolute values of the frequency-dependent complex sheet conductivity σ, in units of milli-Siemens (mS), of the two monolayers and their heterobilayer are presented in Fig. 1(d), where Re(σ) and Im(σ) are the real and imaginary parts, respectively. For extracting the complex sheet conductivity from the experimentally measured data, we used the standard thin-film approximation analysis [32]. The thick continuous curves in Fig. 1(d) are obtained using a polynomial fitting of the data and represent the mean behavior of the conductivity. It can be seen that Re(σ) of all our samples is nearly zero within our experimental accuracy (±0.23 mS) in the entire THz frequency range. The latter uncertainty is mainly related to the uncertainty in the thickness of the quartz substrate (±0.1 µm). For the $MoS_2$ monolayer, Im(σ) is almost zero while it is significantly large for the $MoSe_2$ monolayer. As expected, for the $MoSe_2MoS_2$-heterobilayer, Im(σ) is nearly equal to the geometric mean of the values for the two constituting layers.

***Optical pump THz probe experiment:*** The THz conductivity of the layers can be modulated by modulating their free carrier density. We achieve this using sub-50 femtosecond laser pulses in the visible. The electronic bandgap energies in $MoSe_2$ and $MoS_2$ are ~1.3 eV and 1.9 eV, respectively. Therefore, we use optical pump pulses centered at ~3.1 eV (400 nm) and 1.55 eV (800 nm) for above and below $MoS_2$-bandgap photoexcitation of the $MoSe_2MoS_2$-heterobilayer. The change in transmissivity or reflectivity of the sample sensed by the THz pulse, sampled at its maximum, was recorded versus the time delay between the optical and THz pulses. From this time-resolved photoinduced reflectivity (ΔR(t)) and transmittance (ΔT(t)) of the THz pulses, the modulation in the THz conductivity (Δσ(t)) are directly inferred in real time since Δσ ∝ ΔR or –ΔT [See supplementary information]. Optical excitation at 3.1 eV generates photocarriers in both the monolayer samples and the $MoSe_2MoS_2$-heterobilayer. This modulates the photoinduced conductivity and makes it possible to infer the characteristic hot carriers' relaxation kinetics in them. Hereafter, we will only report on the experiments we performed in reflection mode. A THz echo pulse due to reflection of probe pulse from the back surface of the substrate appears at delay time of ~9ps that gets superimposed on the relaxation curves. To avoid a complex fit procedure, we restricted the analysis of the ΔR signals up to ~9 ps only, however, it is worth mentioning that all the data contain a weak and very slow relaxation components having its time constant ~100 ps (see supplementary information). The corresponding time-resolved ΔR(t) of the THz pulses at various optical pump-fluences are presented in Figs. 2(a), (d) and (g) for the $MoSe_2$ monolayer, $MoS_2$ monolayer and the $MoSe_2MoS_2$-heterobilayer, respectively. Thick curves in Figs. 2(a), (d) and (g) are numerical fits to the data using exponentially decaying functions convolved with a 200-fs Gaussian probe pulse (see supplementary information). Fluence dependence of the kinetic fit parameters, i.e., amplitudes (A's) and time-constants (τ's) of the relaxation components are presented in Figs. 2(b,c) for $MoSe_2$ monolayer, 2(e,f) for $MoS_2$ monolayer and 2(h,i) for the $MoSe_2MoS_2$. Here, the solid curves have been drawn as guide to the eyes and represent the mean behavior of the parameters. There are mainly two relaxation components, a slow one with time constant $\tau_1$ in the range of 10-20 ps whose amplitude, $A_1$ gets saturated quickly at a fluence of ~60 μJ/cm², and another faster one ($A_2,\tau_2$) which becomes significant at higher fluences. The amplitude, $A_2$ of the fast component (time-constant, $\tau_2$ in the range of 0.8-1.2 ps) rapidly increases with the pump-fluence and gets saturated at much higher fluence value of ~150 μJ/cm². Besides the fact that the relaxation becomes faster at higher fluences, we note two more highlights from Fig. 2, (i) $\tau_1$ for $MoSe_2MoS_2$ and $MoS_2$ are very similar, and (ii) at low fluences the magnitude of ΔR for $MoSe_2MoS_2$ is at least 3 times larger than that in any of the two layers.

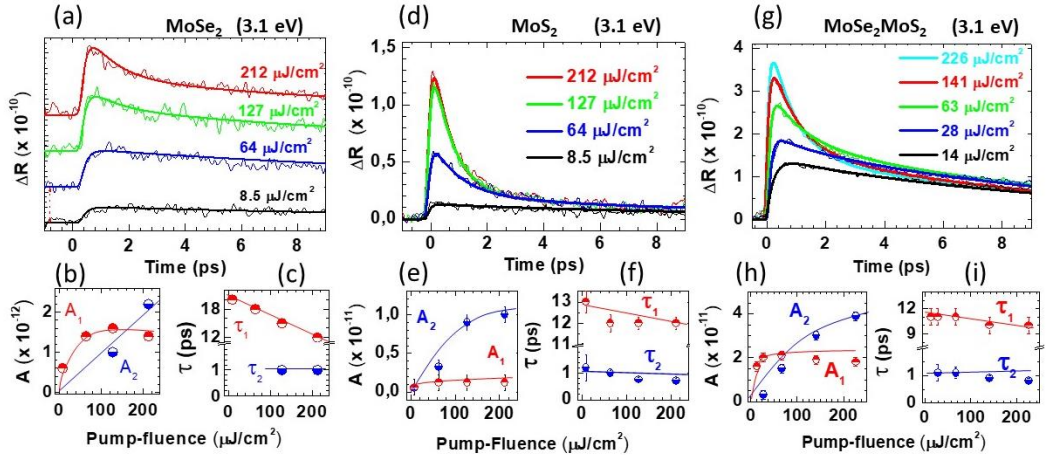

**FIGURE 2.** Optical pump-induced transient THz reflectivity data and pump fluence dependence of the kinetic parameters recorded using optical pump at 400 nm (3.1 eV). Continuous curves in the time-resolved spectra are fits using either single exponential decay or two exponential decay components convolved with a Gaussian pulse (200 fs FWHM). A's are amplitudes and τ's are time-constants of the exponentially decaying components. Time-resolved data at various pump fluences as mentioned in the upper panel and the corresponding exponential fit parameters as a function of the pump fluence in the lower two panels obtained for **(a-c)** $MoSe_2$, **(d-f)** $MoS_2$ and **(g-i)** $MoSe_2MoS_2$-heterobilayer. Consecutive time-resolved data for $MoSe_2$ in (a) are vertically shifted for better clarity. The continuous curves in the lower two panels for each sample are drawn as a guide to the eyes to represent the mean behavior of the fluence-dependence.

The characteristic time-constants, $\tau_1$ ~ 16 ps and $\tau_2$ ~ 1 ps in $MoSe_2$ (Figs. 2(a-c)) are similar to those reported in the literature, for example, for liquid phase exfoliated $MoSe_2$ obtained using transient absorption spectroscopy [33]. Similarly, the two time-constant for $MoS_2$ in Figs. 2(d-f) also match with those from the literature [18], obtained using transient optical absorption spectroscopy. At high fluences, the appearance of the fast time constant $\tau_2$ in all three samples (Fig. 2) is attributed to the exciton-exciton annihilation due to highly injected initial exciton density [34]. Immediately after generation of photocarriers, formation of excitons occurs naturally in these direct band gap TMD monolayers, which after a few ps, decay by emitting photoluminescence [35].

The case with optical excitation at 1.55 eV is even more interesting. In absence of photocarriers, there was absolutely no photoinduced ΔR, i.e., photoinduced THz conductivity in $MoS_2$ up to fluences as high as 500 μJ/cm² (see Fig. S8 in the supplementary information). On the other hand, significant changes in ΔR(t) are obtained for $MoSe_2$ and $MoSe_2MoS_2$, as presented in Figs. 3(a) and (d) with the corresponding kinetic fit parameters shown in Figs. 3(b,c) and 3(e,f), respectively. Here also, a positive change in the photoinduced THz reflectivity, ΔR is obtained for both the $MoSe_2$ monolayer and the $MoSe_2MoS_2$-heterobilayer. Like before, the thick curves in Figs. 3(a,d) are numerical fits and dashed curves in Figs. 3(b,c) and 3(e,f) represent the mean behavior of the fluence dependence. The highlights from these results are: (i) only the slow relaxation component, $(A_1,\tau_1)$ is present up to reasonably high pump-fluences whose time constant strongly decreases with the increasing fluence, while, the saturation in its amplitude is seen at a much higher value of ~150 µJ/cm$^2$, for both the MoSe$_2$ and MoSe$_2$MoS$_2$, (ii) at any given pump-fluence, the magnitude of photoinduced THz reflectivity change in MoSe$_2$MoS$_2$ is about four times stronger than in MoSe$_2$.

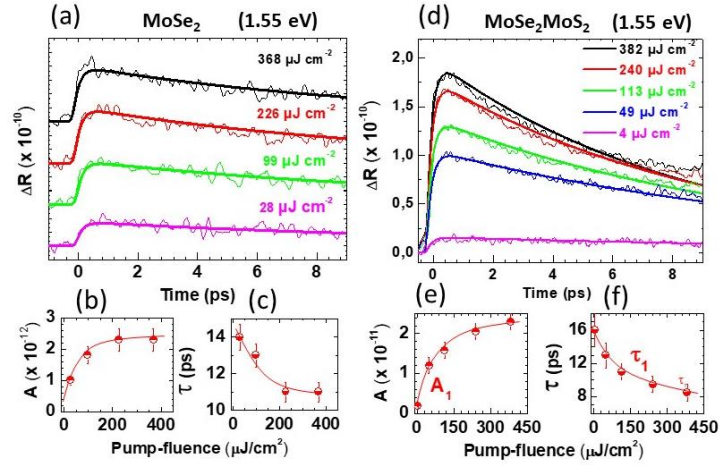

**FIGURE 3.** Optical pump-induced transient THz reflectivity data and pump fluence dependence of the kinetic parameters recorded using optical pump at 800 nm (1.55 eV). Continuous curves in the time-resolved spectra are fits using either single or double exponential decay components convoluted with a Gaussian pulse (200 fs FWHM). A's are amplitudes and τ's are time-constants of the exponentially decaying components. Time-resolved data at various pump fluences as mentioned in the upper panel and the corresponding exponential fit parameters as a function of the pump fluence in the lower two panels obtained for **(a-c)** MoSe$_2$, **(d-f)** MoSe$_2$MoS$_2$-heterobilayer. The continuous curves in the lower two panels are drawn as a guide to the eyes. Consecutive time-resolved data for MoSe$_2$ in **(a)** are vertically shifted for better clarity.

***Discussion:*** Consider the pictorial representations of the MoSe$_2$MoS$_2$-heterobilayer in real space (Fig. 4(a)) and energy space (Fig. 4(b)). Optical excitation at 3.1 eV is sufficiently high to create photoelectrons in both the layers while 1.55 eV excitation produces free carriers only in the MoSe$_2$ layer. Instantaneous displacement of the energetic photocarriers in the entire heterostructure, including the interfacial region, should be responsible for the enhanced photoinduced THz response of MoSe$_2$MoS$_2$ at both the excitation photon energies used here. Assuming the out of plane velocity component of the free electrons to be ~10$^5$ m/s and interlayer separation of ~0.7 nm, it takes only about 10 fs for the energetic electrons to switch sides across the vdW barrier in the heterobilayer.

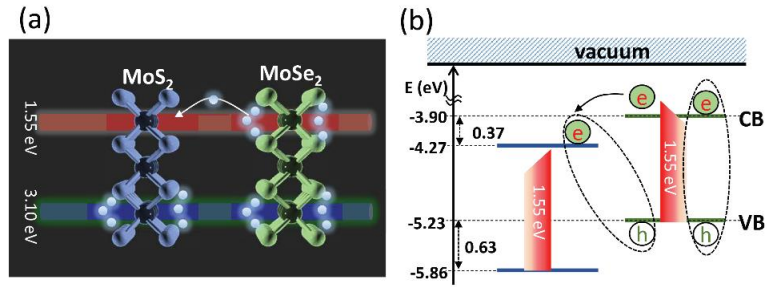

**FIGURE 4.** **(a)** Schematic of the optical excitation at 1.55 eV and 3.1 eV, and **(b)** the processes related to interlayer charge transfer and formation of excitons following optical excitation at 1.55 eV which is above (below) the band gap of MoSe$_2$ (MoS$_2$). Along the vertical axis various energy values from vacuum have been indicated.

The electronic structures of TMD heterostructures have been computed using density functional theory by different groups and consistent numbers for the direct bandgap energy values have been reported [10,36–40]. In the MoSe$_2$MoS$_2$-heterobilayer, the direct bandgap energies of MoSe$_2$ and MoS$_2$ can be taken as ~1.3 eV and ~1.6-1.9 eV, respectively. According to DFT calculations, the number for MoS$_2$ is slightly smaller than what is suggested from optical absorption data [41]. Compared to MoS$_2$, the valence band (VB) and conduction band (CB) of MoSe$_2$ are upshifted by about ~0.63 eV and ~0.37 eV, respectively as indicated in Fig. 4(b). Consequently, like for 3.1 eV, optical excitation of the MoSe$_2$MoS$_2$-heterobilayer at 1.55 eV produces indirect excitons if photoelectrons are transferred from MoSe$_2$ to the MoS$_2$ layer as indicated in Fig. 4(b). In fact, the larger experimental time constant $\tau_1$ in MoSe$_2$MoS$_2$ for 1.55 eV excitation (Fig. 3(f)) than that for 3.1 eV excitation (Fig. 2(i)) suggests a quicker and preferred route to formation of indirect excitons in the heterostructure.

The enhancement in the photoinduced THz response of the MoSe$_2$MoS$_2$-heterobilayer is due to the excited state ultrafast charge transfer between the two layers. This process takes place within the pump pulse duration itself, i.e., much before the intralayer relaxation through direct exciton formation, etc. begins. For the above band gap optical excitation at 3.1 eV, photocarriers are generated in both the layers. They can move across the interface between the two layers of MoSe$_2$MoS$_2$. The movement of the charge carriers continues proportionally up till a certain fluence, i.e., a certain photogenerated carrier density. Beyond this value of the fluence, saturation of the THz response takes place at ~60 µJ/cm$^2$ (Fig. 2(h)). To account for the saturation of the photoinduced THz response of the MoSe$_2$MoS$_2$-heterobilayer, build-up of a reverse electric field $E_r$ has to be considered during the interlayer charge transfer at ultrafast time-scales. At high fluences, band realignment occurs between CBs of MoSe$_2$ and MoS$_2$ (Fig. 4(b)) which stops the further interlayer charge flow. According to this scenario, as soon as the charge transfer stops, the usual exciton formation and carrier recombination within each monolayer takes place. With the carrier recombination, the magnitude of $E_r$ starts to reduce and CBs offset returns to its equilibrium value. In a latest work by Ma et al. [24], generation of THz pulses from above band gap optically excited MoSe$_2$MoS$_2$-heterobilayer was studied. The process responsible for THz generation was suggested to result from the fast interlayer charge transfer in the heterostructure taking place within 100 fs. Moreover, the amplitude of the generated THz pulses was shown to saturate as the pump pulse fluence is increased beyond a certain value such that the CBs offset is nullified temporarily by the excessive fluence.

The above processes can be better visualized from the results obtained at optical excitation of 1.55 eV, i.e., when generation of photocarriers has taken place only in the MoSe$_2$ layer of the heterobilayer (Fig. 4) and photoelectrons' movement is unidirectional towards the MoS$_2$ layer. Since, within the pump-pulse duration, the carriers are moved from MoSe$_2$ to MoS$_2$ conduction band, the saturation fluence at this optical excitation is much larger, about 150 µJ.cm$^{-2}$ (Fig. 3(e)). Upon saturation and nulling of the CBs offset after a while, the movement of the electrons from the MoSe$_2$ to MoS$_2$ stops. The electrons moved across the interface then form indirect excitons with the holes in the MoSe$_2$. A fraction of the photocarriers that still remains in the MoSe$_2$ forms direct excitons and usual intralayer relaxation process for them continues.

At 3.1 eV photoexcitation, part of the electrons can transfer from MoSe$_2$ towards MoS$_2$ and part of the holes from MoS$_2$ towards MoSe$_2$ to form indirect excitons. Saturation of the response is therefore expected to appear at a lower pump-fluence in this case. Moreover, recombination of carriers can now occur in both MoS$_2$ and MoSe$_2$ monolayers. Therefore, to the first approximation, a faster relaxation of the THz response at 3.1 eV optical excitation (Fig. 2(i)) as compared to that at 1.55 eV (Fig. 3(f)) is expected. At 1.55 eV optical excitation, from the absence of a fast relaxation component up to a very high excitation fluence (Fig. 3(d,e)), it is evidently clear that exciton-exciton annihilation does not take place while significant number of photoelectrons in MoSe$_2$ preferably transfer to the MoS$_2$ layer of the heterobilayer to form indirect excitons. The fraction of photoelectrons that transfer from MoSe$_2$ to MoS$_2$ can be estimated from the absorption at the pump photon energy and the corresponding saturation fluence [24]. Since the conduction band offset between MoSe$_2$ and MoS$_2$ is 0.37 eV (Fig. 4(b)), the reverse bias necessary to compensate for this band offset would be $V \geq 0.37$ V. This reverse bias gets developed across the interface instantly within the duration of the excitation pulse and stops the further flow of the photoelectrons. The corresponding built-in electric field would be, $E_{bi} \sim V/d \sim 0.52$ V/nm, where $d \sim 0.7$ nm is the distance between the monolayers. The photoconductivity can be given by [24] $\sigma_d \sim E_{bi}\varepsilon_0\varepsilon_T$, where $\varepsilon_0$ is the free space permittivity and $\varepsilon_T$ is the out-of-plane dielectric constant. Considering $\varepsilon_T = 10$, we find the density of photoelectrons to be of $\sim 2.9 \times 10^{13}$ cm$^{-2}$. The absorption of the heterostructure at 1.55 eV is ~5% and the saturation pump fluence is ~150 µJ/cm$^2$. From this comparison, we find that about 90% of photocarriers are transferred from MoSe$_2$ to MoS$_2$.

In conclusion, we firstly measured the THz conductivity of large area MoS$_2$ and MoSe$_2$ monolayers, and MoSe$_2$MoS$_2$-heterobilayer. Upon photo-excitation above the bandgap of each monolayer, we have shown that compared to the monolayers, the vertical MoSe$_2$MoS$_2$-heterobilayer exhibits manifold enhancement in the THz conductivity change. The relaxation time of the photo-induced carriers are very similar in the single layer and the heterostructure. Moreover, when exciting the heterostructure at 1.55 eV, about 90% of the photocarriers generated in the MoSe$_2$ are almost instantaneously transferred to MoS$_2$. This enhancement and optical control of the photoinduced THz response of MoSe$_2$MoS$_2$-heterobilayer as compared to the monolayers alone should have many implications in future electronic materials engineering for improved performance in optoelectronic applications [33]. Our results indicate that in such a TMD heterostructure an all-optical control of conductivities is possible. This opens the possibility of fast electronic components created for optimal performance at GHz to THz frequencies.

*Acknowledgements:* SK acknowledges SERB, Department of Science and Technology, Govt. of India for financial support through project ECR/2016/000022. IIT Delhi is acknowledged for supporting the research lab, Femtosecond Spectroscopy and Nonlinear Photonics (FemtoSpec Lab). EF acknowledges supports of the Conseil Regional Nouvelle Aquitaine and FEDER for funding the equipment of the COLA platform. YHL acknowledges the financial support by the Institute for Basic Science (IBS-R011-D1).

*Disclosures:* The authors declare no conflicts of interest.

*Supporting information:* This article contains supporting information.

# Supplementary Information

# Enhancement in optically induced ultrafast THz response of MoSe₂MoS₂ heterobilayer

Sunil Kumar,[1,*] Arvind Singh,[1] Sandeep Kumar,[1] Anand Nivedan,[1] Marc Tondusson,[2]
Jerome Degert,[2] Jean Oberle,[2] Seok Joon Yun,[3] Young Hee Lee,[3] and Eric Freysz[2,*]

[1]*Femtosecond Spectroscopy and Nonlinear Photonics Laboratory,
Department of Physics, Indian Institute of Technology Delhi, New Delhi 110016, India*
[2]*Univ. Bordeaux, CNRS, LOMA UMR 5798, 33405 Talence, France*
[3]*Center for Integrated Nanostructure Physics (CINAP),
Institute for Basic Science (IBS), Sungkyunkwan University, Suwon 16419, Republic of Korea*
*eric.freysz@u-bordeaux.fr*
*kumarsunil@physics.iitd.ac.in*

## 1. UV-Visible absorption spectra of the TMD samples

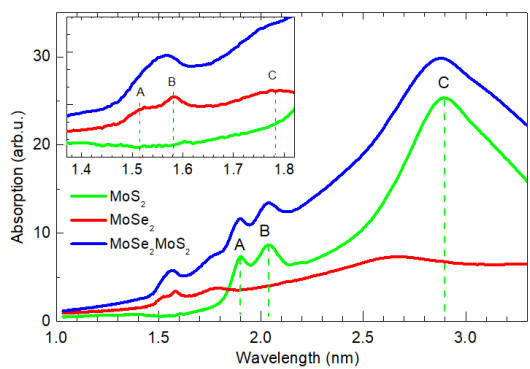

**FIGURE S1.** UV-Visible absorption spectra of the MoS₂, MoSe₂ monolayers and the MoSe₂MoS₂ heterobilayer, all supported on quartz substrates. The main excitonic features displaying the direct band gap nature of the MoS₂ and MoSe₂ monolayers are marked by vertical dashed lines. There is no apparent shift in the excitonic peaks of MoS₂ or the MoSe₂ monolayers due to the presence of the other layer in the heterostructure. The A-, B- and C-excitonic absorption peaks in the MoSe₂ appear at 1.51, 1.58 and 1.79 eV, while those in MoS₂ appear at 1.90, 2.04 and 2.89 eV.

## 2. Raman microscopy of the MoSe₂MoS₂-heterobilayer

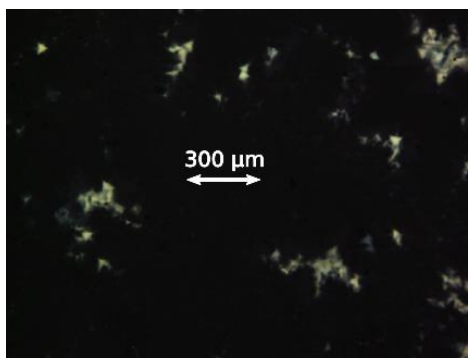

**FIGURE S2.** Raman micrograph of the MoSe₂MoS₂ heterobilayer recorded at the $E_{2g}^1$ mode frequency of the MoSe₂ at 288 cm⁻¹. The black regions correspond to the desired material of the film while the white regions are undesired ones and represent either discontinuities or the MoS₂ monolayers on the substrate. Unambiguously, the desired material covers almost 90% of the surface of the sample. We may note that in the MoSe₂MoS₂-heterobilayer fabrication process, MoSe₂ monolayer was transferred on top of the CVD grown MoS₂ monolayer sample on the quartz substrate.

## 3. THz parameters of the quartz substrate from time-domain spectroscopy

THz electric field waveforms, $E_{THz}(t)$ in air and transmitted through the quartz substrate are shown in Fig. S3(a), while the corresponding spectra, $E_{THz}(\Omega)$ are presented in Fig. S3(b). It can be seen that the substrate absorbs strongly at frequencies beyond ~6 THz. The average THz absorption of the quartz substrate and the three samples, i.e., $MoSe_2$ monolayer, $MoS_2$ monolayer and the $MoSe_2MoS_2$-heterobilayer, can be extracted from the corresponding transmission coefficients provided in Table S1, which were obtained by analyzing the time-domain waveforms given in Fig. 1(c).

**TABLE S1:** THz electric field transmission coefficients (T) obtained from the maximum values of the time-domain THz electric field waveforms. The error bars are indicative of the uncertainty in the thicknesses of the quartz plates.

| S.No. | Sample | Transmission coefficient $T = E_{sample}(t)/E_{air}(t)$ |
|---|---|---|
| 1 | Quartz substrate | $2.64/5.04 = 0.524 \pm 0.003$ |
| 2 | $MoSe_2$ | $(2.274/5.04)/0.524 = 0.861 \pm 0.004$ |
| 3 | $MoS_2$ | $(2.306/5.04)/0.524 = 0.873 \pm 0.004$ |
| 4 | $MoSe_2MoS_2$ | $(2.317/5.04)/0.524 = 0.877 \pm 0.004$ |

Complex refractive index of the substrate, $n_{sub}(\Omega) = n(\Omega) + i\kappa(\Omega)$, n being the index of refraction and $\kappa$ the extinction coefficient, is estimated from the following relation [1],

$$T_{sub}(\Omega) = \frac{4n_{sub}(\Omega)}{(1+n_{sub}(\Omega))^2} e^{-i\Omega d_{sub}\frac{n_{sub}(\Omega)-1}{c}} \tag{S.1}$$

In the above equation, $T_{sub}(\Omega) = E_{sub}(\Omega)/E_{air}(\Omega)$ is the experimental transmission coefficient, $d_{sub}$ is the thickness of the substrate measured independently, c is speed of light in vacuum and $\Omega$ is angular frequency. By comparing the experimental $T(\Omega)$ with the expression in Eq. (S.1), we have extracted $n_{sub}(\Omega)$ and hence its real and imaginary parts as shown in Fig. S3(c).

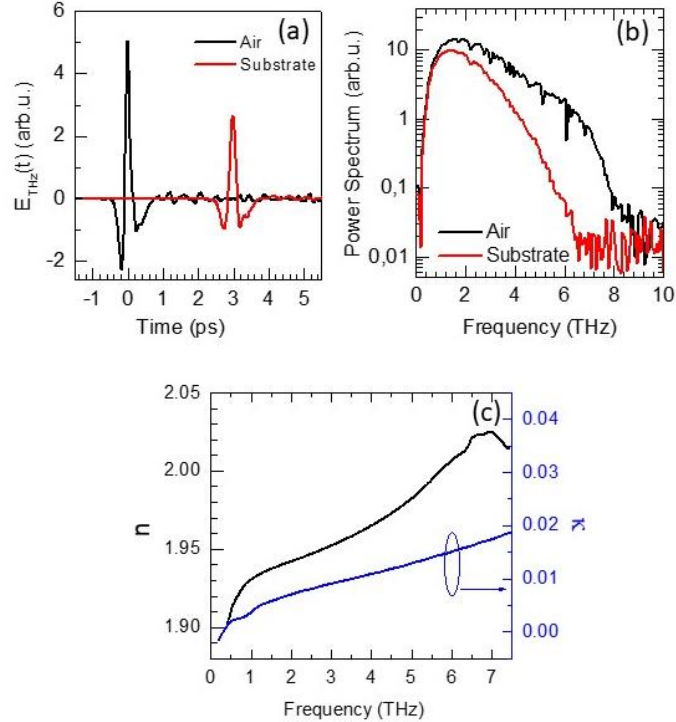

**FIGURE S3.** THz characterization of the quartz plate used as substrate for the TMD layers studied in the paper. (a) Time-domain THz waveforms of the associated complex electric field $E_{THz}(t)$ as recorded without any sample (Air) and with the substrate. (b) Power spectra of the THz signal in air and with the substrate. The allowed THz bandwidth in the experiments is >8 THz, however, due to strong THz absorption in the substrate at frequencies beyond 6 THz, the relevant frequency range is limited to 0.2-5.5 THz. (c) Index of refraction *n* and the extinction coefficient $\kappa$ of the quartz substrate extracted from the complex transmission spectra.

Our experiments performed on another quartz plate from a different batch of the substrates shows slightly different results as shown below in Fig. S4. The seemingly different THz characteristics in the two cases are not surprising because the quartz plates may vary from one batch to another, in their optical properties, more at THz frequencies at which the manufactures do not pay much attention. The results shown in the figure below match well with some of the data available in literature [2]. Of course, FTIR experiments can provide the absorption coefficients at higher frequencies but the THz time-domain spectroscopy provides the simultaneous measurement of the index of refraction also. The quartz plates used as substrates for our thin samples, were all taken from the same batch of substrates and their THz characteristics are same as shown in Fig. S3. It may be noted that this difference in the quartz substrate characteristics has no impact on the measurements and analysis of the properties of thin samples as discussed in our main paper.

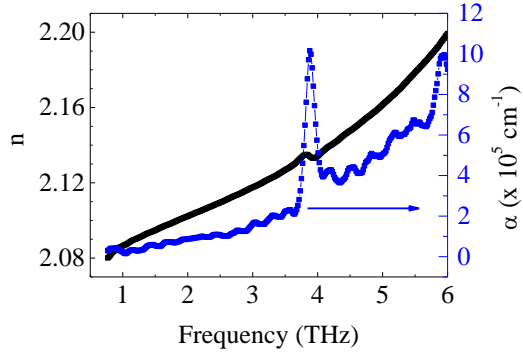

**FIGURE S4.** The index of refraction and the absorption coefficient of a quartz plate taken from a different batch of the substrates. The parameters have been extracted using the same method as describe above in Fig. S3. These results are similar to those reported in the literature [2].

### 4. Data fitting procedure for the optically induced THz response

The temporal evolution of the optical pump induced THz response, i.e., the ΔR and ΔT was analyzed by numerically fitting the experimental data using a convolution function. The latter is described by the following equation,

$$Y(t) = \left(H(t) \sum_i A_i e^{-t/\tau_i}\right) \otimes \exp(-t^2/\tau_p^2) \qquad (S.2)$$

The first function H(t) on the right side of above equation is the standard Heaviside step function to account for material response at only positive times from the instant optical pump excites the sample. The temporal evolution of the material response is captured mainly by the second function, which is a sum of exponentially decaying functions having respective amplitude $A_i$ and time-constant, $\tau_i$. Finally, convolution of the product of the above two functions is taken with the Gaussian probe pulse having time-duration, $\tau_p$ = 200 fs as described in Eq. (S.2) to numerically fit the experimental data in our study using MATLAB.

The fitting procedure to extract the kinetic parameters (amplitudes and time-constants) of the optical pump-induced THz reflectivity and transmittance are described in Figs. S3, S4 and S5.

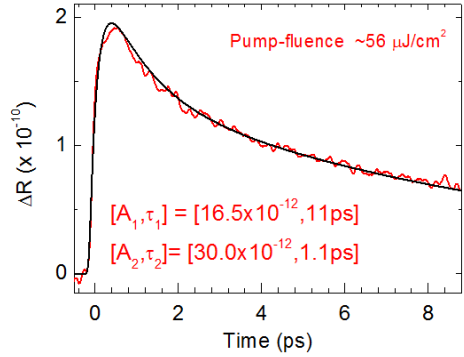

**FIGURE S5:** Transient THz reflectivity (ΔR) of MoSe$_2$MoS$_2$ heterostructure recorded for optical pump wavelength (photon energy) of 400 nm (3.1 eV) shown here to demonstrate the data fitting procedure at a given optical pump-fluence as mentioned. Convolution of a tri-exponentially decaying function with a Gaussian probe pulse having FWHM of 200 fs is used to fit the data. The fit is shown using thick black curve. The major components are [A$_1$,τ$_1$; A$_2$,τ$_2$]: A's are the amplitudes and τ's are corresponding decay time constants while the slowest component [A$_3$,τ$_3$] with decay time-constant > 100 ps is negligibly small.

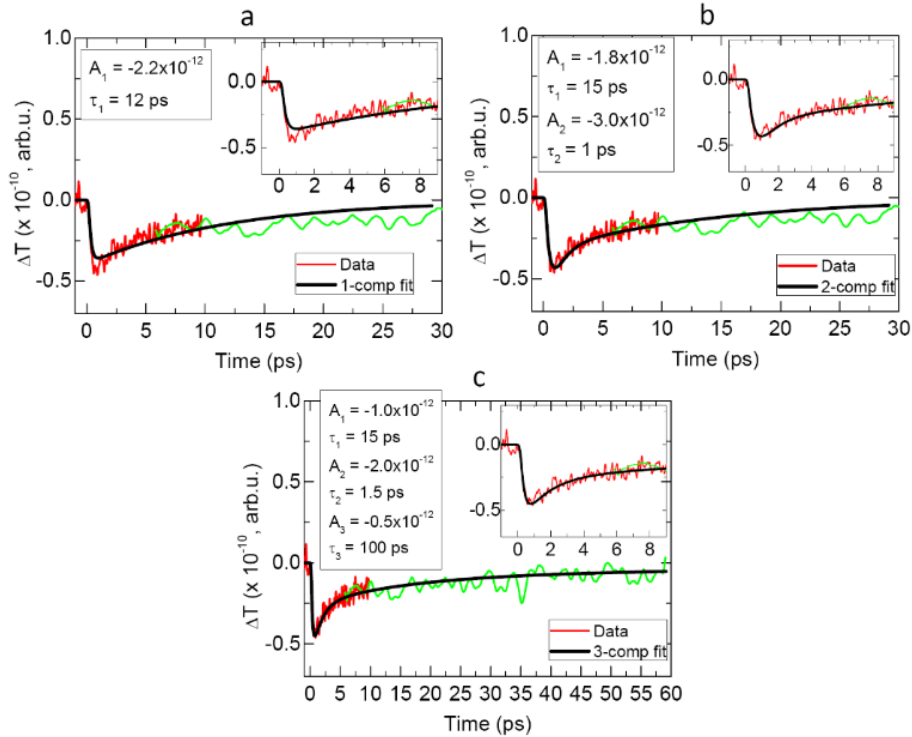

**FIGURE S6.** Comparison between data fitting using one-, two- and three-component exponentially decaying function convolved with the Gaussian probe pulse. The data shown here is for the transient THz transmission change ΔT in MoSe$_2$ at optical pump wavelength (photon energy) of 800 nm (1.5 eV) and pump-fluence of ~215 μJ/cm$^2$. The signals obtained are extremely weak at this pump wavelength. Data with the fits are shown using (a) one-component, (b) two-component, and (c) three-component decaying function in a large time-window. The insets show the same data in narrow time window around the zero delay. Fits in each case have been shown using thick black curves and the corresponding fit parameters have also been mentioned therein. A's are the amplitudes and τ's are the corresponding decay time constants. Clearly, the best fit is obtained for a three-component decaying function having the slowest components being significantly small in comparison with the first two.

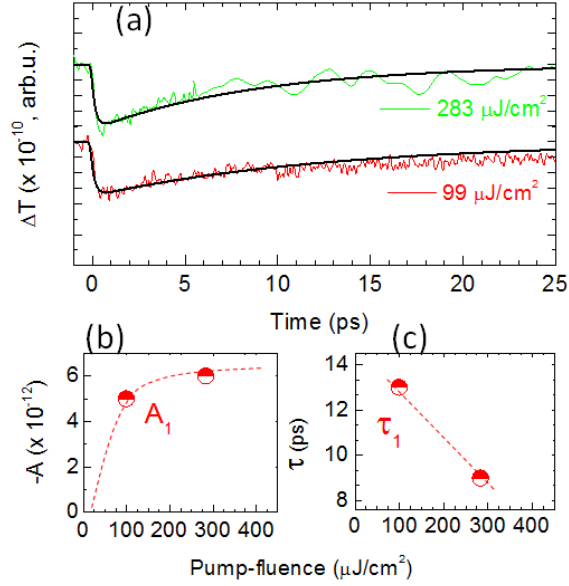

**FIGURE S7.** Optical pump at 800 nm (photon energy 1.5 eV) induced changes in the transient THz probe transmittance signal (ΔT) recorded on $MoSe_2MoS_2$ heterostructure at medium and high pump-fluences as mentioned. The signals are quite weak even under the optimized experimental conditions. The data in (a) can be fitted reasonably well with just single exponentially decaying function having its amplitude $A_1$ and time-constant $\tau_1$ varying with the fluence as shown in the lower panels (b) and (c), respectively. A very weak third component (not shown here) with its time-constant (>100 ps) is required to ensure the best fitting of the data at longer times. We may note that even at such high pump-fluences, the signature of the fast decay constant (~1ps) is absent.

## 5. Optical pump-THz probe experiments performed on $MoS_2$ monolayer at 1.55 eV optical excitation

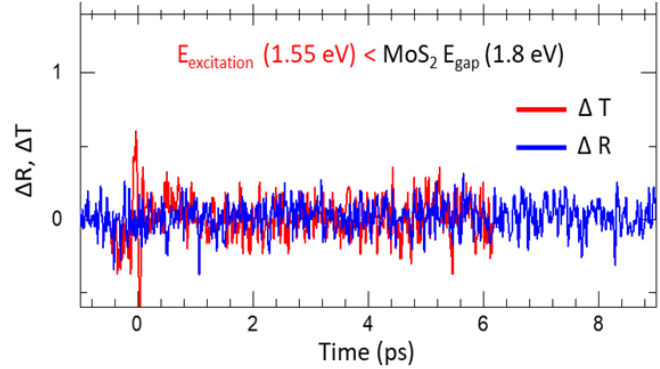

**FIGURE S8.** Optical pump at 800 nm (photon energy 1.5 eV) induced changes in the transient THz probe transmittance (ΔT) and reflectance (ΔR) signals recorded on $MoS_2$ monolayer at high pump-fluence of ~400 μJ/cm².

## 6. THz conductivity from the pump-induced change in THz response

One can roughly estimate the change in the THz conductivity [3] using the following expression valid at normal incidence,

$$\Delta\sigma(\Omega) = \frac{n_s^2 - 1}{2Z_0} \frac{\Delta E(\Omega)}{E(\Omega)} \quad (S.3)$$

Here, $n_s$ and $Z_0$ are the index of the substrate and the impedance of vacuum, respectively. Accordingly, for $MoS_2MoSe_2$ heterostructure at 800 nm, for a fluence of 382 μJ/cm², we find the change in amplitude is ~27%, which corresponds to a change in the sheet conductivity of ~2 mS.

## 7. Sample uniformity

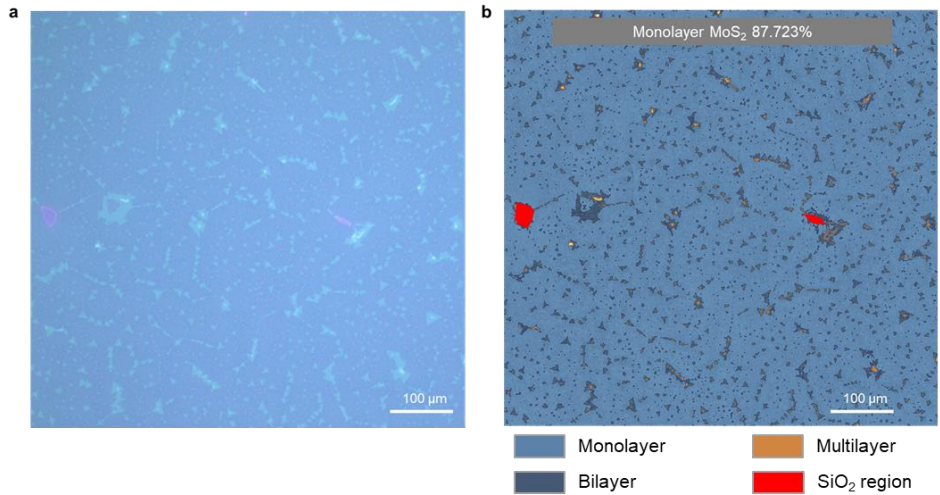

**FIGURE S9.** a. Optical image of CVD-grown MoS$_2$ film sample. b. Monolayer-area marked image by free software of Gwyddion.

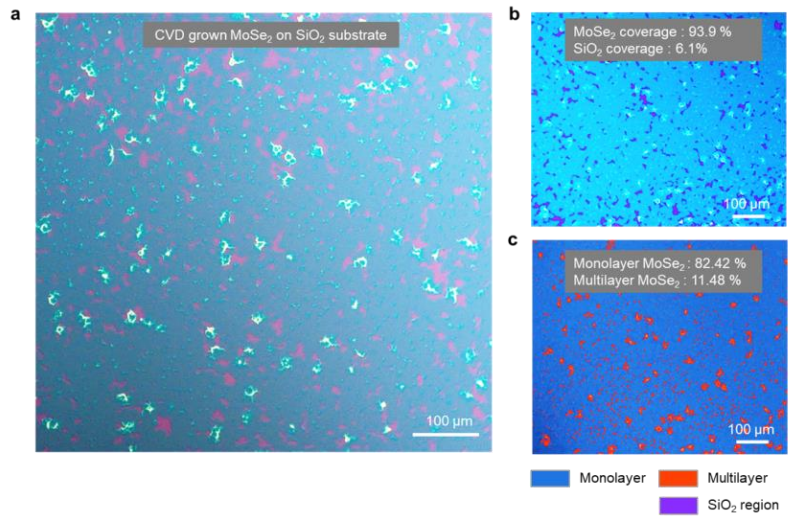

**FIGURE S10.** a. Optical image of CVD-grown MoSe$_2$ sample. MoSe$_2$-area (b) and monolayer MoSe$_2$ (c) marked image by free software of Gwyddion.